\documentclass[11pt]{article}
\usepackage{epsfig} 
\newcommand{\sect}[1]{ \section{#1} \setcounter{equation}{0} }

\newcommand{\half}{\mbox{\small{$\frac{1}{2}$}}} 
\newcommand{\Nc}{N_{\!c}}

\begin{document}
\title{Classification and one loop renormalization of dimension six and eight 
operators in quantum gluodynamics} 
\author{J.A. Gracey, \\ Theoretical Physics Division, \\ Department 
of Mathematical Sciences, \\ University of Liverpool, \\ Peach Street, \\ 
Liverpool, \\ L69 7ZF, \\ United Kingdom.} 
\date{} 
\maketitle 
\vspace{5cm} 
\noindent 
{\bf Abstract.} We determine the complete set of independent dimension six and 
eight Lorentz scalar operators in Yang-Mills theory for an arbitrary colour
group. The anomalous dimension mixing matrix is determined at one loop. 

\vspace{-16cm} 
\hspace{13.5cm} 
{\bf LTH 536} 

\newpage 

\sect{Introduction.} 
Quantum chromodynamics, (QCD), is widely accepted as the quantum field theory
which describes the strong interactions of the nuclear constituents. Indeed at
large energies the theory behaves as if it were virtually a free field theory
allowing one to apply perturbative techniques to describe high energy parton 
interactions. However, a full understanding of the strong interactions at lower
energies scales is still sought. For example, the generation of quark masses 
and quark confinement are not fully understood and are believed to be 
intimately related to the infrared properties of QCD. One approach to 
understand such phenomena is to use effective field theories or models which 
have similar properties to the original QCD Lagrangian. One such model is the 
Nambu-Jona-Lasinio model, \cite{njl}, which involves four quark interaction
terms. From the point of view of standard renormalization theory such terms are
not renormalizable in four space-time dimensions. This is readily apparent from
a simple dimensional analysis since the canonical dimension of such 
interactions is six and therefore their coupling must incorporate a 
dimensionful scale to have a Lagrangian of canonical dimension four. An 
alternative point of view of these operators is that at large energies where
the perturbative approximation is valid, the coupling of these operators is
driven to zero and they are known as irrelevant operators. However, it could be
the case that in the approach to the infrared the coupling, or the anomalous
dimension of the operator, gains a large correction to alter the canonical 
dimension in such a way that they become relevant in the infrared r\'{e}gime.
Therefore, they would then be regarded as sensible and important operators for
understanding the phenomenology at such scales. Whilst such operators and 
models or effective field theories have received wide attention the actual
connection of these models with the original QCD Lagrangian has yet to be 
fully established in detail. Since one can analyse the structure of the gauge 
invariant composite operators of any quantum field theory from the point of 
view of renormalization theory, it is the purpose of this article to consider 
the operators of dimension six which one can be built in Yang-Mills theories. 
The aim is to determine the basis set of dimension six operators and then to 
compute their anomalous dimensions at one loop. 

There are various reasons for such a study aside from those already stated. 
First, we will concentrate on Yang-Mills theories with an arbitrary colour 
group since previous analyses of this problem, we believe, have been 
incomplete. Therefore, it is appropriate to focus on the gluonic sector of QCD 
before returning to the full theory in a later article. For instance, the 
earlier work of \cite{mor} only considered dimension six operators which did 
not involve covariant derivatives of the gluon field strength. We will 
demonstrate that there is an extra independent operator which was omitted from 
that analysis. Second, whilst the work of \cite{sim1,sim2} considered such 
field strength covariant derivatives, the operators were only considered at 
zero momentum. To fully treat composite operators and the determination of 
their anomalous dimensions one must renormalize them with a non-zero momentum 
flowing through them. As is well known doing otherwise can lead to erroneous 
anomalous dimensions. (See, for example, \cite{joglee,jog,col}.) Moreover, we 
will show that one cannot readily drop operators which are total space-time 
derivatives of lower dimensional operators but which are overall dimension six.
These are crucial to preserving identities similar to the Bianchi identities 
which are valid in the classical theory and which must be preserved in the 
quantum theory. This is another reason for concentrating on Yang-Mills theory 
since these technical issues can become over-complicated in QCD. Further, 
previous calculations, \cite{mor,nartar}, were only concerned with specific 
unitary colour groups. We take a more general line here by analysing Yang-Mills
theory with a general Lie group which will allow one, for instance, to 
understand the properties of dimension six operators in the large $N_{\! c}$ 
limit. Another reason for this is that the construction of the basis set of 
operators will not need to appeal to the group tensor properties of a 
particular Lie group. For instance, the totally symmetric Casimir, $d^{abc}$, 
only exists in the group $SU(N_{\! c})$ for $N_{\! c}$ $\geq$ $3$ and does not 
always have a counterpart in other classical or exceptional Lie groups. Also, 
it is possible to consider various group representations in the arbitrary case.
Other motivations for considering dimension six operators come from 
calculations such as \cite{dish,bebiki}. In \cite{sim1,sim2,dish}, for example,
dimension six operators have been shown to be important in hadronic scattering.
In \cite{bebiki} the structure of the mixing matrix of anomalous dimensions 
plays a crucial role in the large order perturbative behaviour of physical 
quantities such as the hadronic decay of the $\tau$ or $e^+ e^-$ annihilation 
into hadrons. Essentially the operator or combination or operators which has 
the dominant eigenanomalous dimension drives the structure of the perturbative 
series at large powers of the strong coupling constant and is related to the 
Borel properties of the series. For other (gluonic) correlation functions the 
dominant eigenoperator could be different and therefore it is important to have
the anomalous dimensions for the full set of dimension six operators. Finally, 
although we have focused on the motivation for considering dimension six 
operators, we will also study dimension eight operators at the same level. 
Again previous analyses in our view have not been fully complete and therefore 
it is important to establish the full picture. 

The paper is organised as follows. In section two, we discuss at length the 
background requirements and results for constructing the basis set of dimension
six operators. This is repeated in section three for the case of dimension 
eight operators before discussing the one loop renormalization of all operators
in section four. Concluding remarks are given in section five. 

\sect{Classification of dimension $6$ operators.} 
To systematically classify dimension six and eight gauge {\em invariant}
operators it is appropriate to choose a notation where this important property 
is manifest. As we are dealing with Yang-Mills operators it seems appropriate 
therefore to choose a group valued field strength, $G_{\mu\nu}$, and gauge
potential, $A_\mu$, where
\begin{equation} 
G_{\mu\nu} ~=~ G^a_{\mu\nu} T^a ~~~,~~~ A_\mu ~=~ A^a_\mu T^a 
\end{equation} 
and $T^a$ are the usual colour group generators obeying the Lie algebra of the
colour group $G$ 
\begin{equation}
\left[ T^a , T^b \right] ~=~ i f^{abc} T^c 
\label{liealg} 
\end{equation} 
with structure constants $f^{abc}$ which satisfy the usual Jacobi identity 
\begin{equation} 
f^{abe}f^{cde} ~+~ f^{ace}f^{dbe} ~+~ f^{ade}f^{bce} ~=~ 0 ~.  
\end{equation} 
Moreover, the covariant derivative of a group valued object $X$, satisfies  
\begin{equation}
D_\mu X ~=~ \partial_\mu X ~+~ ig [ A_\mu , X ] 
\end{equation}
where $g$ is the coupling constant. Consequently, one has
\begin{equation}
\left[ D_\mu , D_\nu \right] X ~=~ ig \left[ G_{\mu\nu} , X \right]
\label{Dcomm}
\end{equation} 
where the field strength is defined as 
\begin{equation} 
G_{\mu\nu} ~=~ \partial_\mu A_\nu ~-~ \partial_\nu A_\mu ~+~ 
i g A_\mu \wedge A_\nu ~.  
\end{equation} 
Given the geometric nature of $G_{\mu\nu}$ it satisfies the Bianchi identity 
\begin{equation} 
D_\mu G_{\nu\sigma} ~+~ D_\nu G_{\sigma\mu} ~+~ D_\sigma G_{\mu\nu} ~=~ 0 
\label{bianchi} 
\end{equation} 
which is a basic symmetry property of the Yang-Mills field. In this notation,
the gauge transformations of the following entities are  
\begin{eqnarray}
A_\mu & \rightarrow & U A_\mu U^\dagger ~+~ \frac{i}{g} \left( \partial_\mu U
\right) U^\dagger \nonumber \\ 
G_{\mu\nu} & \rightarrow & U G_{\mu\nu} U^\dagger \nonumber \\
D_\mu G_{\nu\sigma} & \rightarrow & U D_\mu G_{\nu\sigma} U^\dagger 
\label{gaugetrans} 
\end{eqnarray} 
where $U$ is a group valued $x$-dependent unitary matrix with 
\begin{equation} 
U U^\dagger ~=~ U^\dagger U ~=~ 1 ~.  
\label{unitarity} 
\end{equation} 
Hence, to construct gauge invariant operators whether Lorentz scalar or
otherwise one need only consider colour group traces of objects which transform
covariantly under (\ref{gaugetrans}). For Yang-Mills theories such objects 
would be the field strength itself and any number of covariant derivatives of 
it. Objects with a gauge potential clearly cannot lead to a gauge invariant 
operator. To illustrate how one systematically constructs a set of linearly 
independent operators of a specific dimension, which satisfy the criteria of 
the renormalization theorems, we consider dimension six operators in detail 
first. Since operators can be related to each other by operators which involve 
the equation of motion, then we note at the outset that in quantum 
gluodynamics, the equation of motion is  
\begin{equation} 
D_\mu G^{\mu\nu} ~=~ 0 ~.  
\label{eqnm} 
\end{equation} 
Thus if an operator can be related by the symmetries of the theory to another
operator plus one which includes the object $D_\mu G^{\mu\nu}$ then they are
regarded as being dependent, \cite{col}. The method we have followed is first 
to write down all possible structures involving the objects $G_{\mu\nu}$, 
$D_\mu$ and $\partial_\mu$ which are of dimension six, Lorentz scalars and 
which are gauge invariant by the above construction. With this basic set it is 
straightforward to decorate all the slots for the Lorentz indices in all 
possible ways, though initially dropping those operators involving
$D_\mu G^{\mu\nu}$. The ordinary derivative, $\partial_\mu$, is allowed in this
construction as it has dimension one but due to the imposition of gauge
invariance it can only appear outside a trace of operators and hence it occurs
in operators with total derivatives. We will discuss their treatment later, as 
they will be important, and concentrate for the moment on non-total derivative
operators. 

Clearly to have a colour singlet operator one must have at least two field 
strengths in the trace which for dimension six operators implies only one
trace operation is allowed. Therefore, the three allowed structures are  
\begin{equation} 
\mbox{Tr} \left( G_{..} G_{..} G_{..} \right) ~~,~~  
\mbox{Tr} \left( D_{.} G_{..} D_{.} G_{..} \right) ~~,~~  
\mbox{Tr} \left( G_{..} D_{.} D_{.} G_{..} \right) 
\label{sixset}
\end{equation} 
where the dots indicate the location of the Lorentz indices and the covariant
derivative in any string acts {\em only} on the object immediately to the 
right. Although one can simply enumerate all the cases, making use of the 
symmetry properties reduces the amount of work. For instance, a common object 
in the structures is 
\begin{equation} 
D_{\mu_1} \ldots D_{\mu_n} G^{\nu_1 \nu_2} ~.  
\end{equation} 
However, in certain cases the following index pattern will be present 
\begin{equation} 
D_{\mu_1} \ldots D_\mu \ldots D_{\mu_n} G^{\mu \nu_2} ~~~,~~~ 
D_{\mu_1} \ldots D_\mu \ldots D^\mu \ldots D_{\mu_n} G^{\nu_1 \nu_2} 
\label{opcontr}
\end{equation} 
where we have indicated the explicit contractions. (There could be additional
contractions among the sets $\{\mu_1,\ldots ,\mu_n\}$ and $\{\nu_1 , \nu_2 \}$
but we focus on those illustrated explicitly.) For the first example, applying
(\ref{Dcomm}) recursively to move the covariant derivative to the right yields
operators with a higher number of legs together with the operator
\begin{equation}
D_{\mu_1} \ldots D_{\mu_n} D_\mu G^{\mu\nu_2} 
\label{Deomop} 
\end{equation}
which has the same number of legs as the original operator. We define the 
number of legs on an operator as the lowest number of gluon fields in any term 
when the field strength and the covariant derivatives are written in terms of 
$A_\mu$. Clearly, (\ref{Deomop}) is an operator which vanishes on an equation 
of motion. Therefore, the operator we began with can be written in terms of
higher leg operators plus an operator which is ignored from the point of view
of establishing linear independence, \cite{col}. From focusing on this 
particular structure embedded within an operator an algorithm to determine the 
basis set of operators follows naturally. In other words using symmetries we 
write as far as possible the lower leg operators in terms of higher leg ones 
and equation of motion operators. These then become dependent and can be 
dropped from the original set of potential operators. For the second operator 
of (\ref{opcontr}), it can be written in terms of the first operator by 
commuting the covariant derivative $D^\mu$ closest to the field strength to 
obtain higher leg operators plus
\begin{equation}
D_{\mu_1} \ldots D_\mu \ldots D_{\mu_n} D^\mu G^{\nu_1 \nu_2} ~. 
\end{equation}
Using (\ref{bianchi}) this is related to 
\begin{equation} 
D_{\mu_1} \ldots D_\mu \ldots D_{\mu_n} D^{\nu_1} G^{\mu \nu_2} ~+~ 
D_{\mu_1} \ldots D_\mu \ldots D_{\mu_n} D^{\nu_2} G^{\nu_1 \mu} 
\end{equation} 
which is of the form of the first operator of (\ref{opcontr}). Hence any 
operator involving factors of the form of (\ref{opcontr}) where two Lorentz
indices are contracted in a string, are not independent and depend on higher
leg operators. This is a general result which is not limited to dimension six 
operators. From (\ref{sixset}) it is easy to see that no member of the last set
of operators is independent. For the remaining two possibilities in 
(\ref{sixset}) the lemma implies there can be no index contraction in 
$D_{.} G_{..}$ leaving the cases 
\begin{equation}
\mbox{Tr} \left( D_\mu G_{\nu\sigma} D^\mu G^{\nu\sigma} \right) ~~,~~  
\mbox{Tr} \left( D_\mu G_{\nu\sigma} D^\nu G^{\mu\sigma} \right) ~~,~~  
\mbox{Tr} \left( D_\mu G_{\nu\sigma} D^\sigma G^{\mu\nu} \right) ~.
\end{equation}
The last two are related by the antisymmetry of $G_{\mu\nu}$ and either can
be written using the Bianchi identity as proportional to the first which 
therefore means 
\begin{equation} 
{\cal O}_{62} ~=~ \mbox{Tr} \left( D_\mu G_{\nu\sigma} D^\mu G^{\nu\sigma} 
\right) 
\label{op62def} 
\end{equation} 
remains as the only two leg independent dimension six operator. For the 
remaining structure of (\ref{sixset}) there is only one way of slotting the 
Lorentz indices non-trivially giving the independent three leg operator
\begin{equation}
{\cal O}_{61} ~=~ \mbox{Tr} \left( G_{\mu\nu} G^{\nu\sigma} G_\sigma^{~\mu} 
\right) ~. 
\end{equation}
By taking the trace explicitly this is related to  
\begin{equation}
f^{abc} G^a_{~\mu\nu} G^{b \, \nu\sigma} G^{c \,~ \mu}_{~ \sigma} 
\end{equation}
where the term involving the symmetric tensor $\mbox{Tr}\left( T^{(a} T^b 
T^{c)} \right)$ vanishes by symmetry. 

All that remains are the operators involving a total derivative acting on the
trace of two field strength operators. One might expect that such operators are
unimportant as they would not contribute when inserted in diagrams at zero
momentum. However, it will turn out that they are crucial for ensuring 
consistency in the one loop renormalization and preserving identities which 
follow from the Bianchi identity or symmetries. Indeed total derivative
operators are known to be important in QCD. For example, the renormalization
of the axial vector anomaly involves the operator $G^{\mu\nu} 
\tilde{G}_{\mu\nu}$ where $\tilde{G}_{\mu\nu}$ $=$ 
$\epsilon_{\mu\nu\sigma\rho}G^{\sigma\rho}$ and $\epsilon_{\mu\nu\sigma\rho}$
is the usual totally antisymmetric rank four pseudotensor. As is well known  
$G^{\mu\nu} \tilde{G}_{\mu\nu}$ is the total derivative of the Chern Simons
current. To compute the anomalous dimension of the singlet axial current
correctly one must ensure the axial anomaly equation is satisfied as an 
operator equation quantum mechanically. This requires the renormalization of
the total derivative operator $G^{\mu\nu} \tilde{G}_{\mu\nu}$. (See, for 
example, \cite{kod,lar}.) Therefore, in the context of our dimension six and 
eight classification we also consider such operators, though only those which 
are Lorentz scalar and not pseudoscalar. In addition to the earlier lemmas, we 
now introduce new results which relate various operators by a total derivative 
operator. For example, using the distributivity property of $D_\mu$ acting on 
the product of typical objects $X$ and $Y$ which involve products of 
$G_{\mu\nu}$ and its covariant derivatives, we have
\begin{equation}
\mbox{Tr} \left( X D_\mu Y \right) ~=~ \partial_\mu \mbox{Tr} \left( XY 
\right) ~-~ \mbox{Tr} \left( D_\mu X Y \right) 
\label{totdid1} 
\end{equation} 
which implies 
\begin{equation}
\mbox{Tr} \left( X D_\mu X \right) ~=~ \frac{1}{2} \partial_\mu \mbox{Tr} 
\left( XX \right) ~.  
\label{totdid2} 
\end{equation} 
In the first of these results it is tempting to omit the first term on the 
right. However, from the renormalization theorems it, like $G^{\mu\nu} 
\tilde{G}_{\mu\nu}$, can have an anomalous dimension when renormalized at 
non-zero momentum. Moreover, the same relation provides us with the strategy 
for classifying total derivative operators. In addition to beginning with these
operators with the lowest number of legs, one considers those with the lowest 
number of external derivatives and rearranges them to produce ones with a 
higher number. Two basic structures for dimension six emerge, which are
\begin{equation}  
\partial_{.} \mbox{Tr} \left( G_{..} D_{.} G_{..} \right) ~~~,~~~ 
\partial_{.} \partial_{.} \mbox{Tr} \left( G_{..} G_{..} \right) ~.  
\end{equation} 
Ignoring operators which involve the equation of motion, this gives the
following candidate operators, 
\begin{equation} 
\partial^\mu \mbox{Tr} \left( G^{\nu\sigma} D_\mu G_{\nu\sigma} \right) ~~,~~
\partial^\mu \mbox{Tr} \left( G^{\nu\sigma} D_\nu G_{\mu\sigma} \right) ~~,~~
\partial^\mu \partial_\mu \mbox{Tr} \left( G^{\nu\sigma} G_{\nu\sigma} 
\right) ~~,~~
\partial^\mu \partial^\nu \mbox{Tr} \left( G_{\mu\sigma} G^\sigma_{~\nu} 
\right) ~.  
\label{totset} 
\end{equation} 
The first operator is related to the third by (\ref{totdid2}) and is thus not 
independent. Likewise the second operator is related to the first through the
Bianchi identity. Finally, using the results (\ref{totdid1}), (\ref{totdid2}) 
and the Bianchi identity, we have
\begin{equation}
\partial^\mu \mbox{Tr} \left( G_{\mu\nu} D^\sigma G^\nu_{~\sigma} \right) ~=~ 
\partial^\mu \partial^\sigma \mbox{Tr} \left( G_{\mu\nu} G^\nu_{~\sigma} 
\right) ~+~ \frac{1}{4} \partial_\mu \partial^\mu \mbox{Tr} \left( 
G^{\nu\sigma} G_{\nu\sigma} \right) 
\end{equation} 
leaving the last operator, say, as the only independent dimension six one 
from the set of total derivative operators, (\ref{totset}), where only the
results (\ref{totdid1}) and (\ref{totdid2}) were applied. It remains to check 
what relations emerge when these latter results are applied to the set of 
operators which do not initially involve a total derivative. Therefore, if we
consider $\mbox{Tr} \left( D_\mu G_{\nu\sigma} D^\mu G^{\nu\sigma} \right)$ and
apply (\ref{totdid1}) and (\ref{totdid2}) then we find,  
\begin{equation}  
\mbox{Tr} \left( D_\mu G_{\nu\sigma} D^\mu G^{\nu\sigma} \right) ~=~ 
\frac{1}{2} \partial^\mu \partial_\mu \mbox{Tr} \left( G_{\nu\sigma} 
G^{\nu\sigma} \right) ~+~ 2\mbox{Tr} \left( G_{\nu\sigma} D^\nu D_\mu 
G^{\sigma\mu} \right) ~-~ 4 ig \mbox{Tr} \left( G_{\mu\nu} G^{\nu\sigma} 
G_\sigma^{~\mu} \right) ~. 
\end{equation}  
Hence, overall we are left with only two independent dimension six operators. 

\sect{Classification of dimension $8$ operators.} 
The procedure to classify operators of dimension eight follows that for the
dimension six case and rather than reproduce similar arguments we will 
concentrate on the essential differences. First, with the higher dimension it
is clear that more structures akin to (\ref{sixset}) are possible. Moreover, 
one has to consider operators built out of more than one colour group trace. 
Considering the two leg operators first, like the dimension six case there is 
only one such  operator,  
\begin{equation} 
{\cal O}_{821} ~=~ \mbox{Tr} \left[ \left( D_\mu D_\nu G_{\sigma\rho} \right)  
\left( D^\mu D^\nu G^{\sigma\rho} \right) \right] 
\label{op821} 
\end{equation} 
which is a natural generalization of (\ref{op62def}). All other two leg 
operators either involve an equation of motion operator or can be written as 
(\ref{op821}) and higher leg operators. For three leg operators there is a 
similar reduction in the number of possibilities though one is left with  
\begin{eqnarray} 
{\cal O}_{831} &=& \mbox{Tr} \left( D_\mu G_{\nu\sigma} D^\rho G^{\mu\nu} 
G_\rho^{~\sigma} \right) \nonumber \\ 
{\cal O}_{832} &=& \mbox{Tr} \left( D_\mu G_{\nu\sigma} D^\mu G_\rho^{~\nu} 
G^{\rho\sigma} \right) 
\label{op83} 
\end{eqnarray}
as the two independent operators. As there is now a trace over three group 
generators it might be expected that the Feynman rule of each operator will 
involve $d^{abc}$. For the {\em latter} operator the antisymmetry of 
$G_{\mu\nu}$ ensures that only $f^{abc}$ emerges. Whilst using the Bianchi 
identity on the first operator produces 
\begin{equation} 
-~ \frac{1}{2} \mbox{Tr} \left( D_\sigma G_{\mu\nu} D_\rho G^{\mu\nu} 
G^{\sigma\rho} \right) 
\end{equation}
which likewise only involves $f^{abc}$. For operators of the form 
$\mbox{Tr} \left( D_{.} D_{.} G_{..} G_{..} G_{..} \right)$ it might be 
expected that our lemma could still allow for several independent operators. In
other words if there are no contracted indices on the field strength with two 
covariant derivatives then the lemma is not applicable. This leaves the four 
cases
\begin{eqnarray} 
&& \mbox{Tr} \left( D_\rho D_\mu G_{\nu\sigma} G^{\mu\nu} G^{\sigma\rho} 
\right) ~~,~~ 
\mbox{Tr} \left( D_\rho D_\mu G_{\nu\sigma} G^{\mu\rho} G^{\nu\sigma} \right) 
{}~~,~~ \nonumber \\ 
&& \mbox{Tr} \left( D_\rho D_\mu G_{\nu\sigma} G^{\sigma\rho} G^{\mu\nu} 
\right) ~~,~~ 
\mbox{Tr} \left( D_\rho D_\mu G_{\nu\sigma} G^{\nu\sigma} G^{\mu\rho} 
\right) ~. 
\end{eqnarray} 
Clearly the second and fourth operators of this set are each related to four
leg operators. For the remaining two using the Bianchi identity allows one to
rewrite each as either the second or fourth operator. Similar arguments
systematically applied to the other possible structures leave (\ref{op83}) as
the only three leg dimension eight operators. Finally, for the four leg 
operators which therefore involve four field strength factors and no covariant
derivatives, one has the possibility of a double group trace structure. 
Systematically enumerating the allowed Lorentz structures simply leaves the 
eight basic operators  
\begin{eqnarray} 
{\cal O}_{841} &=& \mbox{Tr} \left( G_{\mu\nu} G^{\mu\nu} G_{\sigma\rho} 
G^{\sigma\rho} \right) \nonumber \\ 
{\cal O}_{842} &=& \mbox{Tr} \left( G_{\mu\nu} G_{\sigma\rho} G^{\mu\nu} 
G^{\sigma\rho} \right) \nonumber \\ 
{\cal O}_{843} &=& \mbox{Tr} \left( G_{\mu\nu} G^{\nu\sigma} G_{\sigma\rho} 
G^{\rho\mu} \right) \nonumber \\ 
{\cal O}_{844} &=& \mbox{Tr} \left( G_{\mu\nu} G_{\sigma\rho} G^{\nu\sigma} 
G^{\rho\mu} \right) \nonumber \\ 
{\cal O}_{845} &=& \mbox{Tr} \left( G_{\mu\nu} G^{\mu\nu} \right) 
\mbox{Tr} \left( G_{\sigma\rho} G^{\sigma\rho} \right) \nonumber \\ 
{\cal O}_{846} &=& \mbox{Tr} \left( G_{\mu\nu} G^{\nu\sigma} \right) 
\mbox{Tr} \left( G_{\sigma\rho} G^{\rho\mu} \right) \nonumber \\ 
{\cal O}_{847} &=& \mbox{Tr} \left( G_{\mu\nu} G_{\sigma\rho} \right) 
\mbox{Tr} \left( G^{\mu\nu} G^{\sigma\rho} \right) \nonumber \\ 
{\cal O}_{848} &=& \mbox{Tr} \left( G_{\mu\nu} G_{\sigma\rho} \right) 
\mbox{Tr} \left( G^{\mu\sigma} G^{\nu\rho} \right) ~.  
\label{op84} 
\end{eqnarray} 

The remaining operators which must be considered now involve those which are
total derivatives. As before if we focus on those operators which have at least
one external derivative acting on the trace of the field strengths then the
possible candidates for independence are, after relating possible operators
within the same structures by, for example, the Bianchi identity, 
\begin{eqnarray}
&& \partial^\rho \partial_\rho \partial^\mu \partial^\nu \mbox{Tr} \left( 
G_{\mu\sigma} G^\sigma_{~\nu} \right) ~~~,~~~ 
\partial^\sigma \partial_\sigma \partial^\rho \partial_\rho \mbox{Tr} \left( 
G_{\mu\nu} G^{\mu\nu} \right) ~~~,~~~ 
\partial^\mu \partial^\nu \mbox{Tr} \left( D_\mu G_{\sigma\rho} D_\nu 
G^{\sigma\rho} \right) \nonumber \\ 
&& \partial^\mu \partial^\nu \mbox{Tr} \left( G_{\mu\sigma} G_{\nu\rho} 
G^{\sigma\rho} \right) ~~~,~~~ 
\partial^\mu \mbox{Tr} \left( G_{\mu\nu} G^{\sigma\rho} D^\nu 
G_{\sigma\rho} \right) ~~~,~~~  
\partial^\mu \partial_\mu \mbox{Tr} \left( G_{\nu\sigma} G^{\sigma\rho} 
G_\rho^{~\nu} \right) ~. 
\end{eqnarray} 
Of this set the last three can be related to operators which do not involve
total derivatives and therefore they are dependent and excluded from the basis.
For example, if one considers ${\cal O}_{831}$ and applies (\ref{totdid1}) 
we have 
\begin{eqnarray}
\mbox{Tr} \left( D_\mu G_{\nu\sigma} D_\rho G^{\mu\nu} G^{\rho\sigma} 
\right) &=& -~ \mbox{Tr} \left( D_\rho D_\mu G_{\nu\sigma} G^{\mu\nu} 
G^{\rho\sigma} \right) ~-~  
\mbox{Tr} \left( D_\mu G_{\nu\sigma} G^{\mu\nu} D_\rho G^{\rho\sigma} \right)
\nonumber \\  
&& -~ \partial^\mu \mbox{Tr} \left( G_{\mu\nu} G^{\sigma\rho} D^\nu 
G_{\sigma\rho} \right) ~-~  
\partial_\mu \mbox{Tr} \left( G^{\sigma\rho} D^\nu G_{\mu\nu} 
G_{\rho\sigma} \right) ~. 
\label{op8totdef} 
\end{eqnarray} 
The first term is related to operators which have already been shown to be
independent by a route not involving (\ref{totdid1}) or (\ref{totdid2}) whilst
the second and fourth terms are equation of motion operators. Repeating the
same manipulations on ${\cal O}_{831}$ but integrating by parts with the other
covariant derivative first, one can relate the fourth operator of 
(\ref{op8totdef}) to the fifth and therefore neither are independent. Next by 
considering ${\cal O}_{832}$ we have  
\begin{equation} 
\mbox{Tr} \left( D_\mu G_{\nu\sigma} D^\mu G^{\rho\nu} G_\rho^{~\sigma} 
\right) ~=~ \frac{1}{6} \partial^\mu \partial_\mu \mbox{Tr} \left( 
G_{\nu\sigma} G^{\rho\nu} G_\rho^{~\sigma} \right) ~-~ \frac{1}{2}  
\mbox{Tr} \left( G_{\nu\sigma} D^\mu D_\mu G_{\rho\nu} G^{\rho\sigma} \right) 
\end{equation} 
where (\ref{totdid1}) has been applied twice. Therefore, to summarize the three
independent dimension eight total derivative operators are 
\begin{eqnarray} 
{\cal O}_{821t} &=& \partial^\rho \partial_\rho \partial^\mu \partial^\nu 
\mbox{Tr} \left( G_{\mu\sigma} G^\sigma_{~\,\nu} \right) ~~,~~
{\cal O}_{822t} ~=~ \partial^\sigma \partial_\sigma \partial^\rho \partial_\rho 
\mbox{Tr} \left( G_{\mu\nu} G^{\mu\nu} \right) \nonumber \\ 
{\cal O}_{823t} &=& \partial^\mu \partial^\nu \mbox{Tr} \left( D_\mu
G_{\sigma\rho} D_\nu G^{\sigma \rho} \right) ~.
\end{eqnarray} 

Whilst this completes the classification of the operators necessary for
renormalizing all dimension six and eight Yang-Mills operators there are 
several points which still need to be addressed. First, each of the operators 
we have produced only represents one in a tower of such gauge invariant colour
singlet operators. For instance, one can introduce tensor products of group 
generators into each trace without spoiling gauge invariance or altering the 
dimension of the operator. The former property follows from the fact that under
the gauge transformations an operator will contain products of $U$ and 
$U^\dagger$ where $U$ is a group element. If group generators are present the 
cancellation of $U$ factors via (\ref{unitarity}) is obstructed. However, it is
possible to show (infinitesimally) that 
\begin{equation} 
U T^a U^\dagger \otimes U T^a U^\dagger ~=~ T^a \otimes T^a 
\end{equation} 
for all $U$. Moreover,
\begin{equation} 
f^{abc} U T^a U^\dagger \otimes U T^b U^\dagger \otimes U T^a U^\dagger ~=~ 
f^{abc} T^a \otimes T^b \otimes T^c ~.  
\end{equation} 
Hence, one can in principle produce an infinite set of additional operators
from the basic set we have constructed. Examples include 
$\mbox{Tr} \left( T^a T^b G_{\mu\nu} T^a G^{\nu\sigma} T^b G_\sigma^{~\mu} 
\right)$ and $\mbox{Tr} \left( T^a D_\mu G_{\nu\sigma} T^a D^\mu G^{\nu\sigma}
\right)$. However, for the latter one can easily reduce this to 
\begin{equation} 
\left( C_F ~-~ \half C_A \right) \mbox{Tr} \left( T^a D_\mu G_{\nu\sigma} T^a 
D^\mu G^{\nu\sigma} \right) 
\end{equation} 
using $T^a T^a$ $=$ $C_F$ where $C_F$ is the usual rank two Casimir and $C_A$ 
is its value in the adjoint representation. For the former the manipulation of 
the group generators would lead to other group Casimirs, \cite{vrscvela}. 
Although this increases the number of possible operators to consider for an 
arbitrary gauge group, we will not classify them but regard them as derivable 
from the base set. It is worth noting that in the explicit renormalization each
additional operator will in fact be generated. This is because inserting them 
in the Green's function group generators are introduced at each vertex through 
the use of the Lie algebra, (\ref{liealg}). However, the main reason we do not 
need to consider such operator generalizations is that at least for the 
classical Lie groups the tensor product $T^a \otimes T^a$ can, in principle, be
decomposed by a group identity. For instance, in $SU(N_{\! c})$ 
\begin{equation} 
T^a_{IJ} T^a_{KL} ~=~ \frac{1}{2} \left[ \delta_{IL} \delta_{JK} ~-~ 
\frac{1}{\Nc} \delta_{IJ} \delta_{KL} \right] ~.  
\label{suncfierz} 
\end{equation} 
The other classical Lie groups have similar relations which allows one to
rewrite operators with generator strings in terms of the base set of operators.
However, the relations are not necessarily related in a group invariant way.
In other words one would have to consider each set of (classical) Lie groups 
separately. As we are interested in performing a calculation without reference 
to particular groups we will allow for the possibility of these new operators 
being generated at each loop order. Therefore, in this context we need to 
define extra dimension eight operators since these will be generated in our one
loop renormalization. They are 
\begin{eqnarray} 
{\cal O}_{841a} &=& \mbox{Tr} \left( G_{\mu\nu} G^{\mu\nu} T^a G_{\sigma\rho} 
G^{\sigma\rho} T^a \right) \nonumber \\ 
{\cal O}_{842a} &=& \mbox{Tr} \left( G_{\mu\nu} G_{\sigma\rho} T^a 
G^{\sigma\rho} G^{\mu\nu} T^a \right) \nonumber \\ 
{\cal O}_{843a} &=& \mbox{Tr} \left( G_{\mu\nu} G^{\nu\sigma} T^a 
G_{\sigma\rho} G^{\rho\mu} T^a \right) \nonumber \\ 
{\cal O}_{844a} &=& \mbox{Tr} \left( G_{\mu\nu} G_{\sigma\rho} T^a 
G^{\nu\sigma} G^{\rho\mu} T^a \right) 
\end{eqnarray} 
and their relation to (\ref{op84}) is readily determined for $G$ $=$ $SU(\Nc)$
using (\ref{suncfierz}), for example.  

One final consideration needs to be addressed and that is the classification
of the dimension six and eight operators which vanish on the equation of 
motion. This is for an important reason. We will renormalize the operators
by inserting them in a Green's function where the external legs are multiplied
by the physical polarization vectors. However, not only will operators which we
have classified be generated but operators which vanish on the equation of 
motion. Not all such operators have a zero Feynman rule when the external legs 
are put on-shell and therefore they can occur with a non-zero pole with respect
to the regularization. Hence, one has to classify such operators and produce a 
basis of them. Once one has determined the set of operators which a specific 
operator mixes into, the explicit mixing matrix of anomalous dimensions of 
{\em physical} operators is determined by excluding the equation of motion 
operators. Given this technical point we have also classified all dimension six
and eight operators which involve the equation of motion (\ref{eqnm}). In this 
construction we proceed as before writing down all possible Lorentz structures 
and using the symmetries to remove dependent operators. However, unlike 
previously we insist that the structures all contain a factor of the form 
$D_\mu G^{\mu\nu}$. This means that for dimension six there are only two leg 
operators and dimension eight at most three leg operators. As the procedure 
then parallels our previous construction we merely write down the operators 
which form the basis for this sector. For dimension six we have 
\begin{equation} 
\mbox{Tr} \left( D_\sigma G^{\nu\sigma} D^\mu G_{\mu\nu} \right) ~~~,~~~ 
\mbox{Tr} \left( G^\nu_{~\sigma} D^\sigma D^\mu G_{\mu\nu} \right)  
\end{equation} 
and for dimension eight the following form the basis 
\begin{eqnarray} 
&& \mbox{Tr} \left( G_{\sigma\rho} D^\nu G^{\sigma\rho} D^\mu G_{\mu\nu} 
\right) ~~~,~~~ 
\mbox{Tr} \left( G^\nu_{~\sigma} G^{\sigma\rho} D_\rho D^\mu G_{\mu\nu} 
\right) ~~~,~~~ 
\mbox{Tr} \left( D^\mu G_{\mu\nu} D^\rho D_\sigma D_\rho G^{\nu\sigma} 
\right) \nonumber \\ 
&& \mbox{Tr} \left( D^\sigma D^\mu G_{\mu\nu} D^\rho D^\nu G_{\sigma\rho} 
\right) ~~~,~~~ 
\mbox{Tr} \left( D^\sigma D^\mu G_{\mu\nu} D_\rho D^\sigma G^{\nu\rho} 
\right) ~~~,~~~ 
\mbox{Tr} \left( D_\sigma G_{\nu\rho} D^\sigma D^\rho D_\mu G^{\mu\nu} 
\right) \nonumber \\  
&& \mbox{Tr} \left( G_{\nu\sigma} D^\sigma D^\rho D_\rho D_\mu G^{\mu\nu} 
\right) ~.  
\end{eqnarray} 
It is worth noting that whilst these are the full set, some but not all vanish 
for all legs when put in a Green's function where the external legs are 
on-shell which therefore reduces the number one has to consider for an operator
to mix into.  

\sect{One loop renormalization.} 
Having established the sets of independent dimension six and eight operators 
which forms the basis, we can now determine their anomalous dimensions. This is
achieved by respecting the standard renormalization theorems, 
\cite{joglee,jog,col}. In essence each operator is inserted in a Green's 
function with the same number of external legs as the operator itself. For 
gauge invariant operators this translates into the lowest number of legs in the
operator since the covariant derivative and gluon field strength involve terms 
with various numbers of legs. When the operator in inserted it has a non-zero 
external momentum flowing into it. If it is inserted at zero momentum one 
cannot readily resolve the resulting set of operators it mixes into 
straightforwardly. Indeed if one considers the simple example of the 
renormalization of $(G_{\mu\nu}^a)^2$ as discussed in \cite{col}, an incorrect 
value for the $\beta$-function would be obtained if one took the naive values 
for the renormalization which emerged. More detailed discussion of this issue
has been given in \cite{hamvnv,fpfrs,cshs}. Therefore, we will insert each 
operator with a non-zero momentum which means that whilst each Green's function
has $n$ gluon legs external, it is in fact an $(n+1)$-point function due to the
extra external momentum. For operators which are a total derivative this 
property is important. The general structure of the Green's function is 
illustrated in figure $1$. In addition to inserting at non-zero
\begin{figure}[ht] 
\epsfig{file=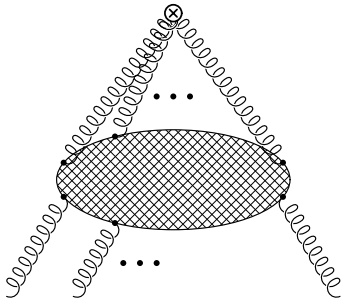,height=4cm}
\vspace{0.5cm}
\caption{Gluonic Green's function with operator insertion.}
\end{figure} 
momentum we exclude the possibility of gauge variant operators emerging in the
mixing matrix by ensuring the external gluons are on-shell. Thus for the 
external gluon $\tilde{A}^a_\mu (p)$ in momentum space, we multiply the Green's 
function by $\epsilon_\mu (p)$, where $\epsilon_\mu$ are the spin-$1$ physical
polarization vectors, and set $p_\mu p^\mu$ $=$ $0$ and $\epsilon_\mu (p) 
p^\mu$ $=$ $0$ for each such gluon. Consequently, when each operator is 
inserted the resulting output will involve a large number of terms involving 
different combinations of the factors $\epsilon_i \epsilon_j$, $p_i \epsilon_j$
($i$ $\neq$ $j$) and $p_i p_j$ ($i$ $\neq$ $j$) where $1$ $\leq$ $i$,$j$ $\leq$
$n$. To resolve these into the operator basis we have computed the Feynman 
rules for each independent operator with the same number of gluon legs as the 
original operator and multiplied by $\epsilon_\mu(p)$ subjecting them to the 
same restrictions as above. The full set with arbitrary coefficients is 
compared with the operator output and the parameters fixed to cancel off all 
the one loop divergences. Given that Yang-Mills is renormalizable and that we 
have a basis set of operators then there is no redundancy or overconstraint in 
the computation of the parameters which are therefore uniquely determined.   

To find the pole structure of each Green's function we have chosen to compute
using dimensional regularization where the space-time dimension is $d$ $=$ $4$ 
$-$ $2\epsilon$. The ultraviolet infinities will emerge as simple poles in 
$\epsilon$ at one loop. However, since we are working with Green's functions 
which are at least $3$-point, (in the case of a two leg operator insertion), we
cannot use the {\sc Mincer} algorithm, \cite{mincer}, which applies only to 
$2$-point propagator type integrals. Moreover, if one uses massless propagators
to compute, say, $4$-point or higher Green's functions then there is a danger 
of obtaining spurious infrared infinities which in dimensional regularization 
are inseparable from the ultraviolet ones we seek. Instead we are forced to 
infrared regularize our one loop integrals by introducing a mass $m$ in the 
gluon propagator which acts as an infrared cutoff. Recently, a similar approach
has been used in \cite{mismun,chmimu} to systematically compute analogous 
Green's functions for dimension five operators in QCD. Therefore, we use for 
our gluon propagator,  
\begin{equation} 
\frac{1}{(k^2+m^2)} \left[ \eta_{\mu\nu} ~-~ 
(1-\alpha) \frac{k_{\mu}k_{\nu}}{(k^2+m^2)} \right] 
\label{gluonprop} 
\end{equation} 
where $m$ appears {\em naturally} as an infrared regularization. Moreover, we
use a covariant gauge fixing with parameter $\alpha$. To ensure that our
renormalization procedure with such a (gauge symmetry violating) propagator is
valid, we have checked the full one loop renormalization of Yang-Mills using
(\ref{gluonprop}) and a mass independent renormalization scheme 
\cite{grwl,pol}. This is important since the operators we are interested in are
composite and therefore each field present in the operator will be renormalized
requiring the wave function renormalization. For example, 
\begin{equation} 
\mbox{Tr} \left( D_\mu G_{\mbox{\footnotesize{o}}\,\nu\sigma} D^\mu 
G^{\nu\sigma}_{\mbox{\footnotesize{o}}} \right) ~=~ Z_A Z_{DG DG}  
\mbox{Tr} \left( D_\mu G_{\nu\sigma} D^\mu G^{\nu\sigma} \right) 
\label{oprendef} 
\end{equation} 
where $A^\mu_{\hbox{\footnotesize{o}}}$ $=$ $Z_A^{\half} A^\mu$ and the 
subscript, $\hbox{\footnotesize{o}}$, denotes the bare quantity. In this
expression $Z_A$ is the usual gauge dependent gluon wave function 
renormalization and $Z_{DG DG}$ is related to the gauge independent anomalous 
dimension of the particular operator which we seek, if we ignore operator
mixing for the moment. Therefore, by computing with a non-zero $\alpha$ we have
a check which is that the operator renormalization constants which emerge must
be gauge independent. Further, by first renormalizing Yang-Mills at one loop, 
this allows us to check that the Feynman rules we use are consistent. This is 
important since given the nature of the operators we are considering, whose
Feynman rules can involve over four thousand terms\footnote{This number 
represents the number of terms in the four and five legs parts of the 
operator ${\cal O}_{844}$. At one loop the six leg part of this operator is not
required for the anomalous dimension.}, we have used a symbolic manipulation 
approach. The Feynman diagrams are generated with {\sc Qgraf}, \cite{qgraf}, 
and converted into a format recognizable by the language {\sc Form}, 
\cite{form}. We have written a programme to convert the {\sc Qgraf} output into
a typical Feynman integral with propagators and vertices substituted. The group
theory for each graph is performed before the integrals are evaluated. This is
achieved by reducing each diaagram to the corresponding vacuum bubble graph by 
systematically rewriting the propagators using the result, 
\cite{mismun,chmimu},   
\begin{equation}
\frac{1}{((k-p)^2+m^2)} ~=~ \frac{1}{(k^2+m^2)} ~+~ 
\frac{(2kp-p^2)}{(k^2+m^2)((k-p)^2+m^2)} ~.  
\end{equation} 
This relation is {\em exact} but for mass regularized propagators all terms
bar the first involve an external momentum in the numerator. If one 
continually repeats the substitution with the termination rule that $O(p^4)$
terms, say, can be dropped due to Yang-Mills renormalizability, then the
procedure will stop leaving only one loop massive vacuum bubbles which are
easily calculated due to  
\begin{equation} 
\int_k \, \frac{1}{(k^2+m^2)^n} ~=~ \frac{\Gamma(n-\half d)} 
{(4\pi)^{\half d}\Gamma(n)} (m^2)^{\half d - n} 
\end{equation}  
for any $n$ $>$ $0$ where $\int_k$ $=$ $\int d^dk/(2\pi)^d$. We have 
implemented the above algorithm in {\sc Form}, \cite{form}. Finally, for 
operators which mix under the renormalization there will be extra terms in 
relations similar to (\ref{oprendef}). In general we have 
\begin{equation} 
{\cal O}_{\mbox{\footnotesize{o}}\, i} ~=~ Z_{ij} {\cal O}_j
\end{equation}
which gives the mixing matrix of anomalous dimensions  
\begin{equation}
\gamma_{ij}(a) ~=~ \mu \frac{\partial ~}{\partial \mu} \ln Z_{ij} ~.  
\end{equation}  

Having summarized our method and the general formalism, we now record our
results. First, for the two independent dimension six operators there is no
mixing and we find that 
\begin{eqnarray} 
\gamma_{61}(a) &=& -~ \frac{C_A}{2} a ~+~ O(a^2) \nonumber \\ 
\gamma_{62}(a) &=& -~ \frac{11}{3} C_A a ~+~ O(a^2) 
\label{op6dims}
\end{eqnarray}
where $a$ $=$ $g^2/(16\pi^2)$ and the subscript on the anomalous dimension 
corresponds to the appropriate operator. The one loop expression for 
$\gamma_{61}(a)$ has been computed previously in \cite{mor,nartar} and we note 
that we get consistency with both calculations which were performed in the 
background field gauge. For completeness we note that in both instances the 
dimension six operator which was considered were multiplied by powers of $a$. 
Including the respective contributions from the $\beta$-function allows one to 
compare the final expressions for the anomalous dimension of each operator. The
anomalous dimension for ${\cal O}_{62}$ is new and is the same as the one loop 
Yang-Mills $\beta$-function, \cite{klzu,codujo}. 

For the dimension eight operators the full one loop mixing matrix of anomalous 
dimensions partitions into blocks defined by the number of legs on the 
operator. Therefore, we list only the entries in each of the blocks. First, for
the single two leg operator 
\begin{equation} 
\gamma_{821,821}(a) ~=~ -~ \frac{11}{3} C_A a ~+~ O(a^2) ~.  
\end{equation} 
For three legs, we find 
\begin{eqnarray} 
\gamma_{831,831}(a) &=& -~ \frac{7}{6} C_A a ~+~ O(a^2) ~~,~~ 
\gamma_{831,832}(a) ~=~ \frac{1}{3} C_A a ~+~ O(a^2) \nonumber \\ 
\gamma_{832,831}(a) &=& O(a^2) ~~,~~ \gamma_{832,832}(a) ~=~ -~ \frac{1}{2} 
C_A a ~+~ O(a^2) 
\end{eqnarray} 
so that this sub-matrix is triangular at this order. For the four leg dimension
eight operators the mixing matrix is further divided into sectors defined by 
the number of colour traces in the original operator. Thus, for the single 
trace operators we have  
\begin{eqnarray} 
\gamma_{841,841}(a) &=& -~ \left( \frac{25}{3}C_F ~-~ 7 C_A \right) a ~+~ 
O(a^2) ~~,~~ \gamma_{841,842}(a) ~=~ \frac{1}{3} C_A a ~+~ O(a^2) \nonumber \\ 
\gamma_{841,843}(a) &=& -~ \left( 16C_F ~-~ 11 C_A \right) a ~+~ O(a^2)  
\nonumber \\ 
\gamma_{841,844}(a) &=& \left( \frac{52}{3}C_F ~-~ 11 C_A \right) a ~+~ O(a^2)
\nonumber \\ 
\gamma_{841,841a}(a) &=& \frac{23}{3} C_F a ~+~ O(a^2) ~~,~~ 
\gamma_{841,842a}(a) ~=~ \frac{2}{3} C_F a ~+~ O(a^2) \nonumber \\ 
\gamma_{841,843a}(a) &=& 16 C_F a ~+~ O(a^2) ~~,~~ 
\gamma_{841,844a}(a) ~=~ -~ \frac{52}{3} C_F a ~+~ O(a^2) \nonumber \\ 
\gamma_{842,841}(a) &=& \left( \frac{50}{3}C_F ~-~ 8 C_A \right) a ~+~ 
O(a^2) ~~,~~ \gamma_{842,842}(a) ~=~ - \frac{4}{3} C_A a ~+~ O(a^2) 
\nonumber \\ 
\gamma_{842,843}(a) &=& -~ \left( \frac{56}{3}C_F ~-~ \frac{38}{3} C_A 
\right) a ~+~ O(a^2) \nonumber \\ 
\gamma_{842,844}(a) &=& \left( 16 C_F - 10 C_A \right) a ~+~ O(a^2) 
\nonumber \\ 
\gamma_{842,841a}(a) &=& -~ \frac{2}{3} C_F a ~+~ O(a^2) ~~,~~ 
\gamma_{842,842a}(a) ~=~ -~ 16 C_F a ~+~ O(a^2) \nonumber \\ 
\gamma_{842,843a}(a) &=& -~ \frac{56}{3} C_F a ~+~ O(a^2) ~~,~~ 
\gamma_{842,844a}(a) ~=~ -~ 16 C_F a ~+~ O(a^2) \nonumber \\ 
\gamma_{843,841}(a) &=& -~ \left( 12 C_F ~-~ \frac{47}{6} C_A \right) a ~+~ 
O(a^2) ~~,~~ \gamma_{843,842}(a) ~=~ O(a^2) \nonumber \\ 
\gamma_{843,843}(a) &=& \left( \frac{34}{3}C_F ~-~ 6 C_A \right) a ~+~ 
O(a^2) ~~,~~ \gamma_{843,844}(a) ~=~ \left( \frac{14}{3} C_F - 2 C_A \right) 
a ~+~ O(a^2) \nonumber \\ 
\gamma_{843,841a}(a) &=& \frac{23}{3} C_F a ~+~ O(a^2) ~~,~~ 
\gamma_{843,842a}(a) ~=~ \frac{13}{3} C_F a ~+~ O(a^2) \nonumber \\ 
\gamma_{843,843a}(a) &=& -~ \frac{34}{3} C_F a ~+~ O(a^2) ~~,~~ 
\gamma_{843,844a}(a) ~=~ -~ \frac{14}{3} C_F a ~+~ O(a^2) \nonumber \\ 
\gamma_{844,841}(a) &=& \left( 6 C_F ~-~ \frac{19}{6} C_A \right) a ~+~ 
O(a^2) ~~,~~ \gamma_{844,842}(a) ~=~ -~ C_A a ~+~ O(a^2) \nonumber \\ 
\gamma_{844,843}(a) &=& -~ \left( \frac{14}{3}C_F ~-~ \frac{11}{3} C_A \right)
a ~+~ O(a^2) \nonumber \\
\gamma_{844,844}(a) &=& -~ \left( \frac{10}{3} C_F - \frac{13}{3} C_A \right)
a ~+~ O(a^2) \nonumber \\ 
\gamma_{844,841a}(a) &=& \frac{5}{3} C_F a ~+~ O(a^2) ~~,~~ 
\gamma_{844,842a}(a) ~=~ -~ \frac{23}{3} C_F a ~+~ O(a^2) \nonumber \\ 
\gamma_{844,843a}(a) &=& \frac{14}{3} C_F a ~+~ O(a^2) ~~,~~ 
\gamma_{844,844a}(a) ~=~ \frac{10}{3} C_F a ~+~ O(a^2) ~.  
\end{eqnarray} 
To simplify the expressions for the anomalous dimensions of the double colour
trace operators we have introduced the intermediate operators 
\begin{eqnarray} 
{\cal O}_{8410} &=& f^{abe} f^{cde} G^a_{\mu\nu} G^{b \, \nu\sigma} 
G^c_{\sigma\rho} G^{d \, \rho \mu} \nonumber \\  
{\cal O}_{8411} &=& f^{abe} f^{cde} G^a_{\mu\nu} G^b_{\sigma\rho} 
G^{c \, \mu\nu} G^{d \, \sigma \rho} 
\end{eqnarray} 
which are not independent as they are related by
\begin{equation}
{\cal O}_{842} ~=~ {\cal O}_{841} ~-~ \half T(R) {\cal O}_{8411} ~~,~~ 
{\cal O}_{844} ~=~ {\cal O}_{843} ~+~ \half T(R) {\cal O}_{8410} 
\end{equation} 
where $\mbox{Tr} \left( T^a T^b \right)$ $=$ $T(R) \delta^{ab}$. Therefore, we 
have 
\begin{eqnarray} 
\gamma_{845,845}(a) &=& \frac{22}{3} C_A a ~+~ O(a^2) ~~,~~ 
\gamma_{845,846}(a) ~=~ O(a^2) ~~,~~ \gamma_{845,847}(a) ~=~ O(a^2) 
\nonumber \\
\gamma_{845,848}(a) &=& O(a^2) ~~,~~ \gamma_{845,8410}(a) ~=~ -~ 28 C_A a ~+~ 
O(a^2) \nonumber \\
\gamma_{845,8411}(a) &=& -~ 2 C_A a ~+~ O(a^2) \nonumber \\  
\gamma_{846,845}(a) &=& \frac{11}{6} C_A a ~+~ O(a^2) ~~,~~ 
\gamma_{846,846}(a) ~=~ O(a^2) ~~,~~ \gamma_{846,847}(a) ~=~ O(a^2) 
\nonumber \\
\gamma_{846,848}(a) &=& O(a^2) ~~,~~ \gamma_{846,8410}(a) ~=~ \frac{4}{3} C_A 
a ~+~ O(a^2) \nonumber \\ 
\gamma_{846,8411}(a) &=& \frac{11}{3} C_A a ~+~ O(a^2) \nonumber \\  
\gamma_{847,845}(a) &=& -~ \frac{1}{3} C_A a ~+~ O(a^2) ~~,~~ 
\gamma_{847,846}(a) ~=~ \frac{28}{3} C_A a ~+~ O(a^2) \nonumber \\  
\gamma_{847,847}(a) &=& -~ \frac{2}{3} C_A a ~+~ O(a^2) ~~,~~ 
\gamma_{847,848}(a) ~=~ -~ 8 C_A a ~+~ O(a^2) \nonumber \\  
\gamma_{847,8410}(a) &=& -~ \frac{34}{3} C_A a ~+~ O(a^2) ~~,~~ 
\gamma_{847,8411}(a) ~=~ \frac{25}{3} C_A a ~+~ O(a^2) \nonumber \\  
\gamma_{848,845}(a) &=& -~ \frac{1}{6} C_A a ~+~ O(a^2) ~~,~~ 
\gamma_{848,846}(a) ~=~ \frac{14}{3} C_A a ~+~ O(a^2) \nonumber \\  
\gamma_{848,847}(a) &=& -~ 4 C_A a ~+~ O(a^2) ~~,~~ 
\gamma_{848,848}(a) ~=~ \frac{10}{3} C_A a ~+~ O(a^2) \nonumber \\  
\gamma_{848,8410}(a) &=& -~ \frac{2}{3} C_A a ~+~ O(a^2) ~~,~~ 
\gamma_{848,8411}(a) ~=~ \frac{11}{3} C_A a ~+~ O(a^2) ~.  
\end{eqnarray} 
Finally, for the total derivative operators we have 
\begin{equation} 
\gamma_{821t,821t}(a) ~=~ \gamma_{822t,822t}(a) ~=~ \gamma_{823t,823t}(a) ~=~
-~ \frac{11}{3} C_A a ~+~ O(a^2) ~.  
\end{equation} 

Having completed the full one loop renormalization we are now in a position
to examine some of the results used in constructing the initial set of 
independent operators. For instance, the operators which involve a total
derivative were related via the identities (\ref{totdid1}) and (\ref{totdid2}).
By considering ${\cal O}_{62}$ these imply, 
\begin{equation} 
\mbox{Tr} \left( G_{\nu\sigma} D_\mu D^\mu G^{\nu\sigma} \right) ~=~ 
\partial_\mu \mbox{Tr} \left( G_{\nu\sigma} D^\mu G^{\nu\sigma} \right) ~-~ 
\mbox{Tr} \left( D_\mu G_{\nu\sigma} D^\mu G^{\nu\sigma} \right) 
\label{idid6} 
\end{equation} 
which was used in \cite{sim1} but with the first term on the right omitted. 
However, one can compute the one loop renormalization of each operator in this 
result following the procedures we have discussed previously. In particular 
each operator is inserted in a gluon $2$-point function with a non-zero 
momentum flowing through the operator. It turns out that the total derivative 
operator of (\ref{idid6}) has an anomalous dimension which is the same as 
${\cal O}_{62}$ whilst their is no renormalization of the other operator. This 
is consistent since it is related to operators which vanish on the equation of 
motion or which involve higher legs and so has no two leg projection. 
Therefore, it would appear that omitting total derivative operators in 
calculations could lead to erroneous results.  

\sect{Discussion.}
We conclude with various remarks. First, we have computed the one loop 
anomalous dimensions of a set of linearly independent gauge invariant dimension
six and eight operators for an {\em arbitrary} Lie group. Whilst we have 
reproduced results that had been derived previously we believe our calculation 
is more comprehensive since the systematic classification of all operators has 
been performed for the first time and operators which are total derivatives 
have been considered. Their effect cannot be neglected since operators are
renormalized at non-zero momentum. Further, it transpires that there is an 
additional operator of dimension six which appears to have been omitted from 
phenomenological considerations. Second, our study has laid the foundation for 
extending the work in various directions. For instance, with the basis of 
independent physical operators it ought to be possible to renormalize these 
operators at two loops in Yang-Mills theory. Also, given the fact that several 
dimension eight operators at one loop have the same anomalous dimensions it 
would be interesting to see if the degeneracy is lifted at this order. 
Moreover, since there are more independent operators in the basis than would 
previously appear to have been considered it would be worthwhile to extend both
the dimension six and eight bases to QCD when quark fields are included. 
Although at one loop previous analyses would seem to suggest a lack of mixing 
between the quark and gluon sectors the extension to two loops would also be 
worth pursuing since examination of the Feynman diagrams which are generated at
two loops suggest that there will be mixing. In a related context the operators
we have focused on have all been Lorentz scalars. Given the recent interest in 
$CP$-violation and the role dimension six operators play in probing physics 
beyond the standard model, \cite{wein}, it will be important to repeat the one 
loop calculations for Lorentz pseudo-scalar operators. Indeed it would be 
interesting to see if an independent analogue of ${\cal O}_{62}$ exists. We 
hope to return to these issues in future work.  

\vspace{1cm} 
\noindent
{\bf Acknowledgements.} The author thanks Dr D.B. Ali and Prof. L. Dixon for 
useful discussions and the latter for pointing out references \cite{dish}. 



\begin{thebibliography}{99} 
\bibitem{njl} Y. Nambu \& G. Jona-Lasinio, Phys. Rev. {\bf 122} (1961), 345. 
\bibitem{mor} A.Yu. Morozov, Sov. J. Nucl. Phys. {\bf 40} (1984), 505. 
\bibitem{sim1} E.H. Simmons, Phys. Lett. {\bf B226} (1989), 132. 
\bibitem{sim2} E.H. Simmons, Phys. Lett. {\bf B246} (1990), 471. 
\bibitem{joglee} S.D. Joglekar \& B.W. Lee, Ann. Phys. {\bf 97} (1976), 160. 
\bibitem{jog} S.D. Joglekar, Ann. Phys. {\bf 100} (1976), 395;
Ann. Phys. {\bf 108} (1977), 233; 
Ann. Phys. {\bf 109} (1977), 210. 
\bibitem{col} J.C. Collins, {\it Renormalization} (Cambridge University Press,
1984).  
\bibitem{nartar} S. Narison \& R. Tarrach, Phys. Lett. {\bf B125} (1983), 217. 
\bibitem{dish} L.J. Dixon \& Y. Shadmi, Nucl. Phys. {\bf B423} (1994), 3;
Nucl. Phys. {\bf B452} (1995), 724. 
\bibitem{bebiki} M. Beneke, V.M. Braun \& N. Kivel, Phys. Lett. {\bf B404} 
(1997), 315. 
\bibitem{kod} J. Kodaira, Nucl. Phys. {\bf B165} (1980), 129. 
\bibitem{lar} S.A. Larin, Phys. Lett. {\bf B303} (1993), 113. 
\bibitem{vrscvela} T. van Ritbergen, J.A.M. Vermaseren \& S.A. Larin, Phys.
Lett. {\bf B400}, 327; 
T. van Ritbergen, A.N. Schellekens \& J.A.M. Vermaseren, Int. J. Mod. Phys.
{\bf A14} (1999), 41. 
\bibitem{hamvnv} R. Hamberg \& W.L. van Neerven, Nucl. Phys. {\bf B379} (1992),
143. 
\bibitem{fpfrs} W. Furmanski \& R. Petronzio, Phys. Lett. {\bf B97} (1980),
437;  
E.G. Floratos, D.A. Ross \& C.T. Sachrajda, Nucl. Phys. {\bf B152}, (1979),
493. 
\bibitem{cshs} J.C. Collins \& R.J. Scalise, Phys. Rev. {\bf D50} (1994), 4117; 
B.W. Harris \& J. Smith, Phys. Rev. {\bf D51} (1995), 4550. 
\bibitem{mincer} S.G. Gorishny, S.A. Larin, L.R. Surguladze \& F.K. Tkachov, 
Comput. Phys. Commun. {\bf 55} (1989), 381;  
S.A. Larin, F.V. Tkachov \& J.A.M. Vermaseren, ``The Form version of Mincer'', 
NIKHEF-H-91-18.
\bibitem{mismun} M. Misiak \& M. M\"{u}nz, Phys. Lett. {\bf B344} (1995), 308.
\bibitem{chmimu} K.G. Chetyrkin, M. Misiak \& M. M\"{u}nz, Nucl. Phys. {\bf 
B518} (1998), 473.
\bibitem{grwl} D.J. Gross \& F.J. Wilczek, Phys. Rev. Lett. {\bf 30}
(1973), 1343. 
\bibitem{pol} H.D. Politzer, Phys. Rev. Lett. {\bf 30} (1973), 1346.
\bibitem{qgraf} P. Nogueira, J. Comput. Phys. {\bf 105} (1993), 279. 
\bibitem{form} J.A.M. Vermaseren, ``{\sc Form}'' version $2.3$, (CAN Amsterdam, 
1992). 
\bibitem{klzu} H. Kluberg-Stern \& J.B. Zuber, Phys. Rev. {\bf D12} (1975), 
467.
\bibitem{codujo} J.C. Collins, A. Duncan \& S.D. Joglekar, Phys. Rev. {\bf D16}
(1977), 438. 
\bibitem{wein} S. Weinberg, Phys. Rev. Lett. {\bf 63} (1989), 2333. 
\end{thebibliography}
\end{document}